\documentclass[%
superscriptaddress,
 amsmath,amssymb,
 aps,
 prD,
 reprint,
]{revtex4-1}
\usepackage[normalem]{ulem}

\usepackage{graphicx}
\usepackage{adjustbox}
\usepackage{dcolumn}
\usepackage{bm}
\usepackage{hyperref}
\usepackage{xcolor}
\definecolor{darkgreen}{rgb}{0, 0.4, 0} 
\definecolor{midgreen}{rgb}{0.5, 0.8, 0.5}
\definecolor{darkred}{rgb}{0.5, 0, 0}
\definecolor{darkblue}{rgb}{0, 0, 0.5} 

\hypersetup{linktocpage=true, colorlinks=true, citecolor={darkblue}, linkcolor={darkred}, urlcolor={darkblue} }
\hypersetup{bookmarks=true, bookmarksnumbered=true, bookmarksopen=true, bookmarksopenlevel=1 }

\usepackage{booktabs}
\usepackage{ulem}
\usepackage{bbold}
\usepackage{comment}
\usepackage{physics}
\usepackage{url}
\usepackage{mathrsfs}
\usepackage{amsmath}
\usepackage{siunitx}
\usepackage{setspace}
\usepackage{scalerel}

\usepackage{titlesec}

\titleformat{\section}{\raggedright\bfseries\large}{\arabic{section}.}{1em}{}
\titlespacing\section{0pt}{10pt plus 4pt minus 2pt}{6pt plus 2pt minus 0pt} 

\titleformat{\subsection}{\raggedright\bfseries}{\arabic{subsection}}{1em}{}
\titlespacing\subsection{0pt}{10pt plus 4pt minus 2pt}{0pt plus 2pt minus 0pt} 

\makeatletter
\renewcommand*{\fnum@figure}{{\normalfont\bfseries \figurename~\thefigure}}
\renewcommand*{\@caption@fignum@sep}{\textbf{: }}
\makeatother

\newcommand{\old}[1]{\hat{#1}}
\newcommand{\nothing}[1]{#1}

\begin{document}

\newcommand{\physicsand}{Department of Physics and Quantitative Biology Institute, Yale University, New Haven, CT, USA}

\title{On the Analytic Origin of Two Species of Cochlear Eigenmodes}

\author{Asheesh S.\ Momi}%
\affiliation{\physicsand}

\author{Isabella R.\ Graf}%
\affiliation{Developmental Biology Unit, European Molecular Biology Laboratory, 69117 Heidelberg, Germany}
\affiliation{Department of Physics and Astronomy, Heidelberg University, 69120 Heidelberg, Germany}

\author{Michael C.\ Abbott}%
 \email{michael.abbott@yale.edu}
\affiliation{\physicsand}

\author{Benjamin B.\ Machta}%
\email{benjamin.machta@yale.edu}
 \affiliation{\physicsand}

\date{March 17, 2026}

\begin{abstract}
After entering the ear, sound waves propagate as surface waves along the cochlea's basilar membrane. In recent work, we showed numerically that the system supports two types of modes: localized resonant modes, which underpin the modern understanding of cochlear mechanics, and a novel class of spatially extended modes. Here, we develop an analytic framework that explains the emergence of this mode structure. We show that extended modes arise from globally continuous standing-wave solutions, whereas localized modes result from internal resonance requiring matching across a singular point. These results clarify the generic structure of cochlear wave equations.
\end{abstract}

\maketitle


\section*{Introduction}
In mammalian hearing, sounds are processed in the inner ear’s cochlea. This spiral-shaped organ is a fluid-filled cavity partitioned by the basilar membrane (BM).  
Incoming sound waves generate surface waves along the BM that are well described by an inhomogeneous wave equation~\cite{reichenbach2014physics}. Due to the spatially varying BM stiffness, waves of different frequencies peak at different locations in the cochlea, effectively allowing it to function as a spatial frequency analyzer for which waves deposit most of their energy at a frequency-specific location~\cite{lighthill1981energy}. 
This energy localization underlies much of our understanding of cochlear mechanics \cite{Julicher2003,reichenbach2014physics,talmadge1998modeling} and is associated with so-called localized resonant modes.

In recent work \cite{MyPaper}, however, we showed numerically that the behaviour of the cochlea can be broken into two large classes of modes: the aforementioned localized modes and a novel set of spatially extended modes. While localized modes align naturally with conventional views of cochlear function, the origin and potential role of extended modes remain unclear. Since our focus in the previous paper~\cite{MyPaper} was on how hair-cell activity tunes localized modes to the edge of instability while maintaining the stability of extended modes \cite{MyPaper}, the origin and structure of these modes were not examined in detail.

Several unresolved phenomena in hearing may be connected to these extended modes. For example, it remains unclear how hearing works at low frequencies since the BM does not have any resonance below approximately  200\,Hz \cite{recio2017mechanical}. Extended modes occur at frequencies lower than those resonant on the BM \cite{MyPaper}, suggesting a possible role in low-frequency sound processing. Furthermore, these modes occur only at a discrete set of frequencies and,
under certain conditions, can become unstable.
Although the extended modes identified in Ref.~\cite{MyPaper} occur at frequencies that are
too low to directly explain spontaneous otoacoustic emissions,
their discrete structure and potential for instability are reminiscent of
that phenomenon, in which healthy ears emit faint tones at specific frequencies \cite{shera2022whistling}.

To date, evidence for this mode structure has been purely computational. Here, we develop an analytic framework that explains the emergence of these two classes of modes. We show that extended modes arise from globally continuous solutions, whereas localized modes result from internal resonance requiring matching across a singular point. Together, these results clarify the mode structure of inhomogeneous wave equations, such as those that model the cochlea, and provide a foundation for understanding how such systems distribute and localize energy.
\section*{Results}
 We begin with exactly the model in  \cite{MyPaper} written in dimensionless variables. The only length scale in the model is the length of the BM, which is used to rescale distance $\nothing{x}=\old{x}/\old{L}$, and the resonant frequency at the start of the ear is used to rescale time, $\nothing{t}=\old\omega_0\old{t}$. In doing so, the wave equation becomes
\begin{equation}
\label{eq:wave_eq}
\partial_{\nothing{t}}^2\nothing{h}(\nothing{x},\nothing{t})=\partial_{\nothing{x}}^2\nothing{p}(\nothing{x},\nothing{t}).
\end{equation}
Here $\nothing{p}$ and $\nothing{h}$ are the nondimensional pressure and height.

To close Eq. \ref{eq:wave_eq}, we need a mechanical model of the basilar membrane and its surrounding fluid.
As is commonly assumed, we take the relationship to be local in space with an exponentially decaying stiffness term~\cite{MyPaper,talmadge1998modeling,greenwood1990cochlear}
\begin{equation}
\label{Imed_RT}
\nothing{{p}}(x,t)=\nothing{\sigma}\left[\partial_{\nothing{t}}^2\nothing{{h}}(x,t)+\nothing{\xi}\partial_{\nothing{t}}\nothing{{h}(x,t)}+e^{-2\nothing{k}\nothing{x}}\nothing{{h}}(x,t)\right]
\end{equation}
 which can be written in the frequency domain via the acoustic impedance $\nothing{Z}$
\begin{equation}
\label{p_Zh}
\nothing{\tilde{p}}(x,\omega)=\nothing{\sigma}\nothing{Z}_{\text{exp}}(\nothing{x},\nothing{\omega})\nothing{\tilde{h}}(\nothing{x},\nothing{\omega})
\end{equation}
where
\begin{equation}
\label{imped}
Z_{\text{exp}}(x,\omega)=-\nothing{\omega}^2+\nothing{\xi}\nothing{\omega}+e^{-2\nothing{k}\nothing{x}}.
\end{equation}
From this procedure come three dimensionless constants: $\nothing{\xi}$ is the inverse of the maximum Q factor ($Q\propto \nothing{\xi}^{-1}$), $\nothing{k}$ characterizes how much the stiffness decreases along the BM, and $\nothing{\sigma}$ is a dimensionless coupling constant arising from the fluid–membrane interaction. A similar process is followed when dealing with the boundary conditions. 

\begin{figure}
    \centering
\includegraphics[width=1\linewidth]{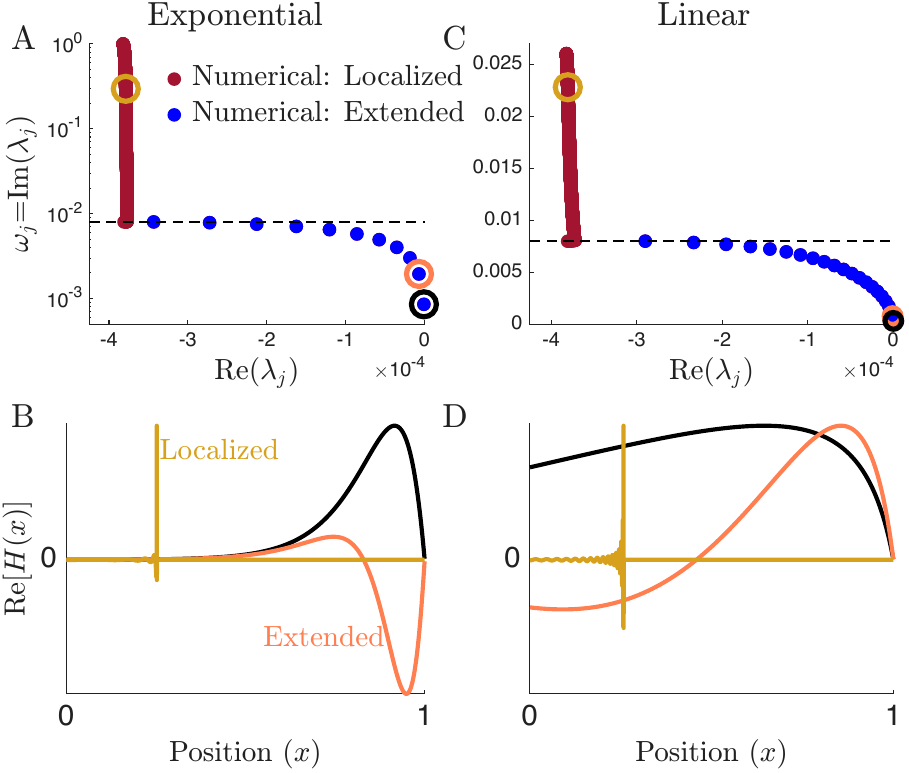}
    \caption{\textbf{A linearly decreasing stiffness is qualitatively similar to an exponential BM.} (A) The eigenvalue structure of cochlear modes \cite{elliott2007state}, with an exponentially decreasing BM stiffness. An eigenvalue's imaginary part determines the oscillation frequency, and the real part determines stability. We define localized modes (red) as those that have a resonant position within the cochlea. Extended modes (blue) are those with frequencies lower than any resonant on the BM. Analogous to \cite{MyPaper} Fig.~2 without active processes. (B) Eigenvectors corresponding to the circled eigenvalues in A. (C) The eigenvalue structure of cochlear modes, with a linearly decreasing BM stiffness. Note that the eigenvalue structure is qualitatively the same, though localized modes are now linearly spread instead of exponentially. The impedance used is from Eq.~\ref{lin_imped} with $x_0=1$  (D) Eigenvectors corresponding to the circled eigenvalues in C. }
    \label{fig:exp/lin}
\end{figure}

\subsection*{Variant Model with Linear Stiffness}
The system of Eqs. \ref{eq:wave_eq}-\ref{imped} has no known analytical solution. 
To better characterize the mode structure analytically, we expand the stiffness in terms of a Taylor series about some point $x_0$ such that the impedance is
\begin{equation}
\label{lin_imped}
    Z(x,\omega)=-\nothing{\omega}^2+i\nothing{\omega}\nothing{\xi} +e^{-2\nothing{k}x_0}[(1-2\nothing{k}(\nothing{x}-x_0)].
\end{equation}
In order for the stiffness to remain positive throughout the BM,  $\nothing{x}_0>1-(2k)^{-1} $. All figures use $x_0=1$.
Fig. \ref{fig:exp/lin} shows that the resulting mode structure is qualitatively preserved, showing a continuum of localized modes and a finite set of extended modes. This justifies examining the modes in a linear cochlea.
Note, however, that the frequency of localized modes now decays linearly instead of exponentially.

\subsection*{Analytic Solutions}

In order to examine the eigenstructure of the systems, it is helpful to write the system in terms of eigenvalues $\lambda$.
This change of variables is done by setting $i\nothing{\omega}=\lambda$ or by using  the ansatz $h(\nothing{x},\nothing{t})=H(\nothing{x})e^{\lambda \nothing{t}}$. 
This changes the impedance to
\begin{equation}
Z(x,\lambda)=\lambda^2+\lambda\nothing{\xi} +e^{-2\nothing{k}x_0}[1-2\nothing{k}(\nothing{x}-x_0)]
\end{equation}
and the wave equation to 
\begin{equation}
\label{lin_wave}
\lambda^2\frac{\nothing{p}(\nothing{x},\lambda)}{\nothing{\sigma} Z(\nothing{x},\lambda)}=\partial^2_{\nothing{x}}\nothing{p}(\nothing{x},\lambda).
\end{equation}
The general solution for the pressure profile can then be written as 
\begin{equation}
\label{solution}
    \begin{aligned}
    \nothing{p}(\nothing{x},\lambda)=& \sqrt{\lambda^2Z(\nothing{x},\lambda)}[c_1I_1( \frac{e^{2\nothing{k}x_0}}{\nothing{k}\nothing{\sigma}}\sqrt{\lambda^2Z(\nothing{x},\lambda)})\\&+c_2K_1( \frac{e^{2\nothing{k}x_0}}{\nothing{k}\nothing{\sigma}}\sqrt{\lambda^2Z(\nothing{x},\lambda)})],
    \end{aligned}
\end{equation}
where $I_1(x),K_1(x)$ are first-order modified Bessel functions of the first and second kind, respectively, and $c_1,c_2$ are both integration constants.

\subsection*{Boundary Conditions for Extended Modes}

\begin{figure}
    \centering
    \includegraphics[width=1\linewidth]{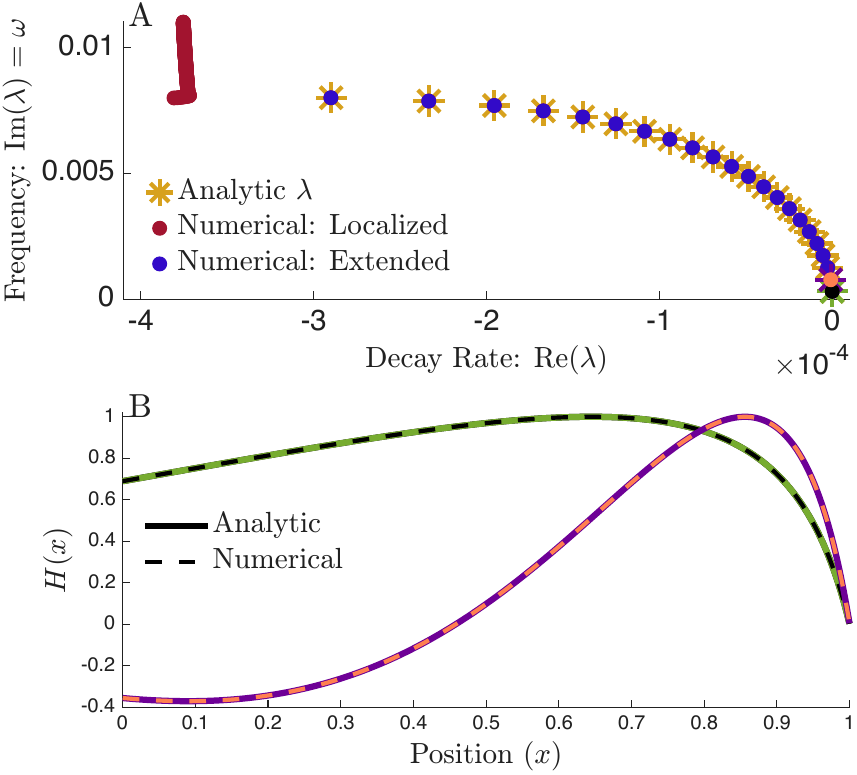}
    \caption{\textbf{Analytic Solution for Extended Modes.}
(A) Eigenvalue structure for a cochlea with linearly decreasing BM stiffness.
Yellow stars denote eigenvalues predicted by the analytic solution
(Eqs.~\ref{solution}-\ref{RHS}), showing excellent agreement with the matrix method. Though the numerical solution is identical to Fig.~\ref{fig:exp/lin} the y-axis is truncated to highlight the extended mode structure. 
(B) Eigenvector corresponding to the marked eigenvalue in (A), demonstrating close agreement between analytic (solid line) and numerical (dashed lines) results.}
    \label{fig:Extended}
\end{figure}

With the analytic solution we can determine the eigenvalues $\lambda$
by imposing the boundary conditions from \cite{talmadge1993new,MyPaper}
\begin{align}
\label{LHS}
 \nothing{\rho}\,\nothing{p}(0,\lambda) &=(\lambda^2+\nothing{\xi}_\text{ow}\lambda+\nothing{\omega}_\text{ow}^2)\frac{1}{\lambda^2}\,\partial_{\nothing{x}} \nothing{p}(x,\lambda)\big\vert_{\nothing{x}=0} 
\\
\label{RHS}
    \nothing{p}(1,\lambda) &= 0.
\end{align}
Here $\nothing{\xi}_\text{ow}$ is the non-dimensionalized friction of the oval window, $\nothing{\omega}_\text{ow}$ is the non-dimensionalized resonant frequency of the oval window, and $\nothing{\rho}$ is an effective coupling constant analogous to $\nothing{\sigma}$. Combining Eq. \ref{solution}  with  these boundary conditions yields a nonlinear equation
$f(\lambda)=0$ whose roots determine the allowed eigenvalues.

Restricting to solutions with spatially continuous eigenfunctions,
we find that all extended modes obtained from the matrix method
are captured by the analytic solution (Fig.~\ref{fig:Extended}A).
Substituting these eigenvalues back into Eq.~\ref{solution} also
produces eigenvectors that closely match the numerical results
(Fig.~\ref{fig:Extended}B).
This agreement confirms that the linear model preserves the eigenvalue structure and spatial form of extended modes while permitting an exact analytic treatment.

The number of extended modes is set by the final localized mode
whose resonant position lies at $\nothing{x}=1$.
If this mode has $M$ zero crossings (22 in Fig.~\ref{fig:Extended}),
then there are $M-1$ extended modes.
Extended modes correspond to continuous global standing-wave
solutions that accumulate nodes as frequency increases.
The $M^{\text{th}}$ mode instead becomes resonant on the BM and transitions into a localized mode.

The value of $M$ is determined by how many wavelengths
fit between $\nothing{x}=0$ and the resonant position at $\nothing{x}=1$.
Parameters such as $\nothing{k}$ and $\nothing{\sigma}$ in the exponential model,
as well as $x_0$ in the linear model, change this accumulated phase
and therefore control the number of extended modes. We observe at least one extended mode so long as the resonant frequency at the end of the ear is underdamped, which in this linear stiffness model means $k<\ln(2/\xi)/2x_0$.

\subsection*{Localized Modes by Matching at Resonance}

\begin{figure}
    \centering
    \includegraphics[width=1\linewidth]{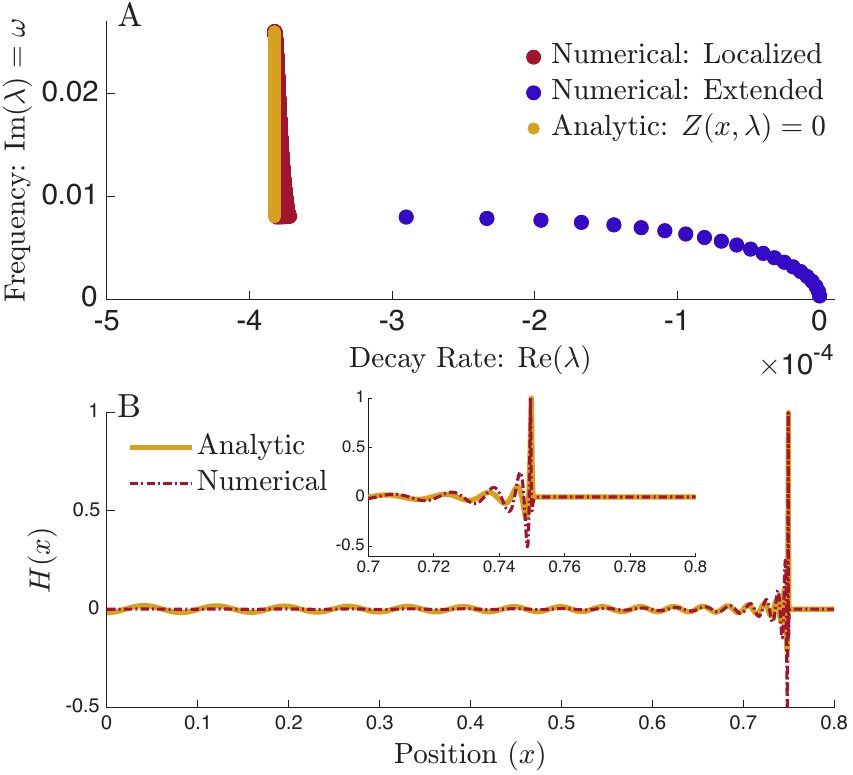}
    \caption{\textbf{Analytic Solution for Localized Modes} (A) ) Localized eigenvalues determined by the resonance condition $Z(x,\lambda)=0$, (yellow line), showing agreement with the numerically obtained localized modes..
 (B) Eigenfunction for a mode resonant at $x=0.75$. The analytic solution obtained by matching the left and right solutions across the resonant point reproduces the spatial structure of the localized mode. The solution is truncated shortly beyond resonance where the amplitude rapidly vanishes.
}
    \label{fig:loc}
\end{figure}

Localized modes require a different treatment than extended modes.
While the analytic solution naturally predicts extended modes,
it fails for localized modes because the impedance vanishes at a resonant position,
$Z(\nothing{x},\lambda)=0$.
This singular point acts as an internal boundary,
requiring the solution to be defined on either side and matched across the resonant position.

As the continuum limit is approached, localized modes approach
the resonance condition $Z(\nothing{x},\lambda)=0$ \cite{MyPaper}.
At this point the modified Bessel function of the second kind,
$K_1(\bar{x})$, diverges,
signaling the breakdown of the continuous solution.

We therefore define localized eigenvalues through the condition
$Z(x_r,\lambda)=0$ and construct solutions on either side of the
resonant position $x_r$. Because the solution is singular at resonance,
Eq.~\ref{solution} is only well defined to the left and right of $x_r$.
Allowing $x_r$ to vary continuously between zero and one therefore
produces a continuum of localized modes.

The pressure profile must therefore take the piecewise form
\begin{equation} 
\label{PW}
\frac{\nothing{p}(\nothing{x},\lambda)}{\sqrt{\lambda^2Z(\nothing{x},\lambda)}}=
 \begin{cases} 
c_{1l}I_1(\bar{x})+c_{2l}K_1(\bar{x}),  &  \nothing{x}< x_r  \\
c_{1r}I_1( \bar{x})+c_{2r}K_1( \bar{x}), &  \nothing{x}> x_r
\end{cases}
\end{equation}
where $\bar{x}= \frac{e^{2\nothing{k}x_0}}{\nothing{k}\nothing{\sigma}}\sqrt{\lambda^2 Z(\nothing{x},\lambda)}$ is used to shorten notation and $c_{1l},c_{1r},c_{2l},c_{2r}$ are all integration constants. Two constants can be fixed with the boundary conditions Eq.~\ref{LHS}-\ref{RHS}
These other two are fixed by enforcing physically motivated
matching conditions at resonance.
First, pressure must remain a continuous function of $x$, including at resonance,
implying $c_{2r}=c_{2l}=c_2$.

The second condition we impose is that the wave equation must hold everywhere, specifically in a small integration window
across the singular point
\begin{equation}
\label{lin_wave}
    \int_{x_r-\epsilon}^{x_r+\epsilon} d\nothing{x}\lambda^2\frac{\nothing{p}(\nothing{x},\lambda)}{\nothing{\sigma} Z(\nothing{x},\lambda)}=    \int_{x_r-\epsilon}^{x_r+\epsilon} d\nothing{x}\partial^2_{\nothing{x}}\nothing{p}(\nothing{x},\lambda).
\end{equation}
The limit where $\epsilon\rightarrow 0$ leads to the condition that $c_{1r}-c_{1l}=i\pi c_{2}$; details are provided in the Appendix. This condition implies a finite jump in the derivative ($\partial_{\nothing{x}}\nothing{p}$),
producing a cusp at the resonant position. 
As shown in Fig.~\ref{fig:loc}, this solution accurately
reproduces the spatial structure of localized modes,
with agreement improving as the discretization approaches
the continuum limit. This demonstrates that localized modes can be understood
as resonance-driven solutions requiring internal matching,
in contrast to the globally continuous extended modes.

\subsection*{WKB Approximation for Exponential Stiffness}

\begin{figure}
    \centering
    \includegraphics[width=\linewidth]{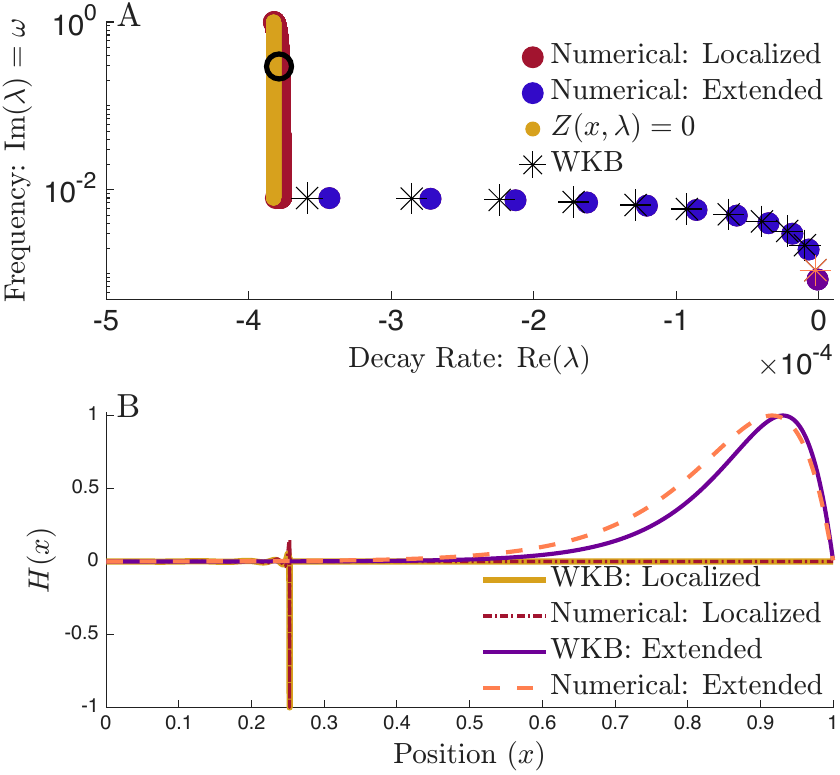}
    \caption{\textbf{Extended and Localized Modes in the WKB Approximation.} (A) The eigenvalue structure of cochlear modes, with a exponentially decreasing BM stiffness. Note that the extended modes predicted by the WKB approximation are qualitatively similar to those obtained numerically. (B) Eigenvector corresponding to the eigenvalues in A with a matching color. Both localized modes are indicated with a black circle.}
    \label{fig:WKB}
\end{figure}
Having obtained analytic insight from the linearized model, we now return to the biologically motivated case of exponentially varying stiffness \cite{greenwood1990cochlear}. For exponentially varying stiffness, an exact analytic solution is not available, so we instead employ the WKB approximation, which assumes a slowly varying impedance and is generally valid away from resonance.
 It has the following form \cite{Julicher2003,talmadge1998modeling,reichenbach2014physics,hulst2025marrying,frost2024foundations}:
\begin{equation}
\nothing{p}_{\scaleto{WKB}{3pt}}(\nothing{x},\lambda)\approx Z(\nothing{x},\lambda)^{\frac{1}{4}}(a_re^{-i\phi(\nothing{x})}+a_le^{i\phi(\nothing{x})}).    
\end{equation}
Here, $\phi(x)=\int_0^xdx'\sqrt{-\lambda^2/Z(x',\lambda)}$ is used to shorten notation and  $a_r,a_l$ are integration constants that multiply the right and left moving basis waves respectively. We can perform the exact same resonance matching discussed above to determine the extended modes; the results are shown in Fig. \ref{fig:WKB}. Although the predicted eigenvalues are not numerically exact,
the WKB approximation captures all qualitative features:
the correct number of modes, the overall trend,
and the structure of the eigenvectors,
which acquire additional zero crossings with increasing frequency.

$Z(x_r,\lambda)=0$ is still used to define the eigenvalues for localized modes. 
However, using the WKB approximation on localized modes eigenfunctions is more subtle, as the approximation breaks down near resonance.

Instead, the matching is governed by the local singular structure,
best explored by expanding $Z$ to linear order as done previously.
We therefore apply the same matching conditions to connect the
left- and right-moving WKB solutions.

To implement the matching conditions, we must relate
$I_1(\bar{x})$ and $K_1(\bar{x})$
to left- and right-moving waves. To do so, we look at the asymptotic behavior as $x\rightarrow\infty$. In this limit, $I_1(\bar{x})\propto e^{i\phi(x)}$ corresponds to a left-moving wave,
while $K_1 (\bar{x})\propto e^{-i\phi(x)}$ corresponds to a right-moving wave.
We therefore define piecewise WKB solutions on either side of resonance
\begin{equation} 
\label{PW_WKB}
\frac{\nothing{p}(\nothing{x},\lambda)}{(\lambda^2Z(\nothing{x},\lambda))^{\frac{1}{4}}}=
 \begin{cases} 
a_{rr}e^{-i\phi(\nothing{x})}+a_{lr}e^{i\phi(\nothing{x})}, &  \nothing{x}< x_r  \\
a_{rl}e^{-i\phi(\nothing{x})}+a_{ll}e^{i\phi(\nothing{x})}, &  \nothing{x}> x_r
\end{cases}
\end{equation}
with the same matching conditions $a_{rr}=a_{rl}=a_r$ and $a_{ll}-a_{lr}=i\pi a_r$. The resulting eigenfunction shows excellent agreement with the numerical solution Fig.~\ref{fig:WKB}. 

\section*{Discussion}
In this work we examined a commonly used inhomogeneous wave equation model of cochlear dynamics and demonstrated that it supports two distinct classes of modes: global extended modes and resonance-driven localized modes.
By combining an exact analytic solution of a linearized model, numerical eigenvalue computations, and a WKB approximation for an exponentially varying stiffness, we clarified the mathematical origin of this mode structure.
Extended modes arise from globally continuous standing-wave solutions, whereas localized modes emerge from internal resonance and require matching across a singular point.

This structure is not unique to the specific model studied here.
Rather, we expect it to be a generic feature of inhomogeneous wave equations in which the local resonant frequency decreases along the membrane while remaining sufficiently high at the apical end.
In such systems, globally extended standing-wave solutions naturally arise in a finite domain. 
At frequencies that match the local resonant frequency, energy is absorbed at the corresponding spatial position, preventing higher-frequency waves from propagating further.
This critical-layer absorption, a general property of cochlear models \cite{lighthill1981energy}, produces resonance-driven localized modes and leads generically to the coexistence of localized and extended modes. In fact, we see this mode structure is preserved when active processes are added \cite{MyPaper}, the functional form of the stiffness and friction is modified \cite{talmadge1998modeling,MyPaper}, and even in two membrane models of cochlear mechanics \cite{neely1986model,MyPaper,elliott2007state}.

In humans, the resonance at the apical end of the BM is 165\,Hz, leaving the mechanism of low-frequency hearing an area of active research \cite{recio2017mechanical,reichenbach2010ratchet}. This frequency range coincides with that of the extended modes identified here,
suggesting that such modes may contribute to low-frequency sound processing.

Spontaneous otoacoustic emissions occur at a discrete set of frequencies though
typically observed at higher frequencies
\cite{talmadge1993new,shera2022whistling}.
Although the extended modes we study lie at lower frequencies,
their discrete structure and instability under certain
tuning conditions \cite{MyPaper}
suggest that related mechanisms could be relevant.

Cubic nonlinearities are a generic feature of cochlear amplification
\cite{hudspeth2010critique,Julicher2003}. Such nonlinearities naturally couple modes of different frequencies,
providing a possible mechanism by which low-frequency extended modes
could generate or influence higher-frequency signatures.
Exploring this nonlinear interaction may clarify whether extended-mode
dynamics contribute to otoacoustic emissions or low frequency hearing.

Together, these results suggest that modal structure plays a
fundamental role in how spatially varying wave systems
organize and redistribute acoustic energy.
\noindent

\begin{acknowledgments}
\noindent
We thank Pascal Martin and members of the Machta
group for helpful discussions and Jose Betancourt for constructive comments on the paper.   This work was supported by NIH R35 GM138341 (BBM).
\end{acknowledgments}

\bibliographystyle{ieeetr}    
\bibliography{bib}

\onecolumngrid

\titleformat{\section}{\raggedright\bfseries\large}{Appendix \arabic{section}.}{1em}{} 
\titleformat{\section}{\setcounter{equation}{0}\raggedright\bfseries\large}{Appendix \Alph{section}.}{1em}{} 
\renewcommand{\thesection}{\Alph{section}}
\setcounter{section}{0}

\renewcommand{\theequation}{\Alph{section}\arabic{equation}}
\setcounter{equation}{0}



\newpage
\begin{center}
\LARGE
\textbf{Appendix}
\end{center}
\normalsize
\noindent
\tableofcontents

\section{Nondimensionalization}
We begin by fully describing the change of variable we did to nondimensionlize the system we described in \cite{MyPaper}. The variables used in \cite{MyPaper} are all dimensionful and denoted with a hat ($\old{x}$) here. To start we take position and normalize it by the  length of the BM and  normalize time using the maximum frequency resonant on the BM so,
\begin{equation} 
\label{xhat}
    \nothing{x}:=\frac{\old{x}}{\old{L}}~~~~~~~~~~~~~~~~\nothing{t}:=\old{t}\old{\omega}_0
\end{equation}
where $\old{L}=3.5$cm, $\old{\omega}_0=2\pi*20800$ rad/s.
Combining these new variables with the original wave equation in \cite{MyPaper} leads to a natural change of variables for pressure and height, 
\begin{align}
\nothing{h}(x,t) &:= k_0\old{h}(\old x, \old t)  & k_0 &:=\sqrt{2\frac{\old \rho \old W_\text{bm}}{\old \sigma_\text{bm}\old A_\text{cs}}} = 3100  \text{m}^{-1} \\
\nothing{p}(x,t) &:=\frac{\old p(\old x, \old t)}{p_0} &p_0 & := \old k_0\old \sigma_\text{bm}\old L^2\old \omega_0^2 = 3.56\times10^9 \text{Pa}
\end{align}
 And then the Fourier transform of these variables becomes, 
 \begin{equation}
    \nothing{\omega}:=\frac{\old{\omega}}{\old \omega_0}~~~~~~~~~~~~~~~~~~~~~~~~~~~~~~\nothing{\tilde{h}}=\old{\tilde{h}} \old k_0 \old\omega_0~~~~~~~~~~~~~~~~~~~~~~~~~~~~~~\nothing{\tilde{p}}=\tilde{p}\frac{\omega_0}{p_0}
\end{equation}
Three dimensionless parameters that come from this procedure are, 
\begin{equation}
    \nothing{\xi}=\frac{\old \xi}{\old \omega_0}=7.65\times 10^{-4}~~~~~~~~~~~~~~~~~~~~~~~~~~~\nothing{k}=\old k \old L=4.83~~~~~~~~~~~~~~~~~~~~~~~~~~~\nothing{\sigma}=(\old k_0 \old L)^{-2}=8.52\times 10^{-5}
\end{equation}
$\nothing{\xi}$ is the inverse of the maximum Q factor $Q=  \nothing{\xi}^{-1}$,
$\nothing{k}$ is the decay rate of stiffness ($e^{-\nothing{k}}$ is the ratio of minimum to maximum resonant frequency on the BM), and 
$\nothing{\sigma}$ is a coupling constant arising from the fluid–BM interaction.
We can then use a similar procedure on the right hand side boundary conditions replacing the displacement of the oval window with,
\begin{equation}
\nothing{d}=\frac{\old d_\text{ow}}{d_0}~~~~~~~~~~~~~~~~~~~~~d_0=\frac{\old L \old W_\text{bm}}{k_0 \old A_\text{ow}}
\end{equation}
which introduces three more dimensionless constants,
\begin{equation}
\nothing{\xi}_\text{ow}=\frac{\old \xi_\text{ow}}{\old \omega_0}=3.83*10^{-3}~~~~~~~~~~~~~~~~~~~~~\nothing{\omega}_\text{ow}=\frac{\old \omega_\text{ow}}{\old \omega_0}=7.21*10^{-2}~~~~~~~~~~~~~~~~~~~~~\nothing{\rho}=\frac{\old A_\text{ow}2 \old \rho \old L }{\old \sigma_\text{ow} \old A_\text{cs}}=11.0
\end{equation}
$\nothing{\xi}_\text{ow}$ is the inverse of the Q factor of the oval window resonating at the start of the BM, 
$\nothing{\omega}_\text{ow}$ is the ratio of oval window resonant frequency to resonant frequency at the start of the BM, and $\nothing{\rho}$ is a coupling constant arising from the fluid–OW interaction analogous to $\nothing{\sigma}$.
\newpage
\section{Finding the Eigenvalues}
We here briefly explain the procedure used to determine the eigenvalues of the system for a fully worked out example see \cite{MyPaper} Appendix A. We start by using a finite element approximation to Eq.~\ref{eq:wave_eq} where we break the BM into $N$ points separated by a distance $\delta \nothing{x}=\frac{1}{N}$. We can write the discretized system in the form,
\begin{equation}
    F\Vec{p}=\partial_{\nothing{t}}^2\Vec{{h}}
\label{vecwave}
\end{equation} 
$\Vec{p}$ is an $N$ column vector with entries here $\vec{p} = \big(\nothing{p}(0,\nothing{t}),\, \nothing{p}(\delta \nothing{x},\nothing{t}),\, \dots,\, \nothing{p}((N-1)\delta \nothing{x},\nothing{t})\big)^\top$, $\Vec{h}$ is an $N$ column vector of BM displacement with the first entry being oval window displacement $\vec{h} = \big(\nothing{d},\, \nothing{h}(\delta \nothing{x},\nothing{t}),\, \dots,\, \nothing{h}((N-1)\delta x,\nothing{t})\big)^\top$, and $\nothing{F}$ is a modified finite difference matrix. We use a modified version of the formalism from Ref. \cite{elliott2007state} to write the dynamics as
\begin{equation}
    \partial_{\nothing{t}}\Vec{X}={J}\Vec{X}
\end{equation}
where ${J}$ is the system's Jacobian and $\Vec{X}$ is the state vector concatenating $\big(\nothing{d},\partial_{\nothing{t}} \nothing{d},\nothing{h}(n\delta \nothing{x},\nothing{t}),\partial_{\nothing{t}} \nothing{h}(n\delta \nothing{x},\nothing{t})\big)$ with $n=1,2,3 ..., N-1$. In order to get the Jacobian of $\vec{X}$ we must first take its time derivative. We can express this time derivative as a sum of a contribution from $\vec{X}$, with the prefactor of each term in $\vec{X}$ given by a matrix $D$, and a contribution from the pressure across the BM $\vec{p}$, with the prefactor of each term in $\vec{p}$ given by a matrix $B$
\begin{equation}
\label{Xdot}
\partial_{\nothing{t}}\vec{X}=D\vec{X}+B\vec{p}.
\end{equation}
Here $D$ is a $2N\times2N$ and $B$ is a $2N\times N$ block diagonal matrix   where entries contain the  information in Eq.~\ref{lin_imped},\ref{LHS}. We can then combine Eq. \ref{vecwave} with Eq. \ref{Xdot} to write,
\begin{equation}
\partial_{\nothing{t}}{\vec{X}}={D}\vec{X}+{B}{F}^{-1}\partial^2_{\nothing{t}}{h}
\end{equation} 
If we then define a $N\times2N$ matrix $E$ such that,
\begin{equation}
\label{Emat}
   \partial_{\nothing{t}} \vec{h}=E\vec{X},
\end{equation}
Then a simple substitution yields
\begin{equation}
\partial_{\nothing{t}}\vec{X}={D}\vec{X}+{B}{F}^{-1}{E}\partial_{\nothing{t}}\vec{X}.
\end{equation}
Finally, a rearrangement yields,
\begin{equation}
\label{fin_jac}
\begin{split}
    \partial_{\nothing{t}}\vec{X}&={J} \vec{X}\\
    {J}&= ( \mathbb{1} -{B}{F}^{-1}{E})^{-1}{D}
    \end{split}
\end{equation}
where ${J}$ is the Jacobian of the system, a $2N\times2N$ matrix.
\newpage
\section{Solving for Extended Modes}
\begin{figure}
    \centering
\includegraphics[width=0.6\linewidth]{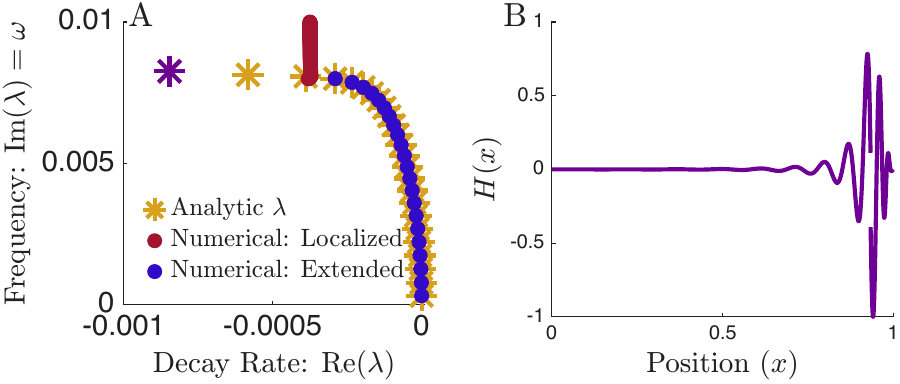}
    \caption{\textbf{Discontinuous Additional Extended Modes.}
(A) This has the same data as Fig.~\ref{fig:Extended} with a large range of x values so we see some additional extended modes predicted left of the red line. 
(B) A demonstration that these additional solutions are invalid as they have discontinuous eigenfunctions.}
\label{fig:discont}
\end{figure}

We begin with the with Eq.~\ref{solution},
\begin{equation}
\label{asolution}
    \begin{aligned}
    p(\nothing{x},\lambda)= \sqrt{\lambda^2Z(\nothing{x},\lambda)}[c_1I_1( \frac{e^{2\nothing{k}x_0}}{\nothing{k}\nothing{\sigma}}\sqrt{\lambda^2Z(\nothing{x},\lambda)})+c_2K_1( \frac{e^{2\nothing{k}x_0}}{\nothing{k}\nothing{\sigma}}\sqrt{\lambda^2Z(\nothing{x},\lambda)})].
    \end{aligned}
\end{equation}
applying the left hand side boundary condition,
\begin{equation}
\label{aLHS}
(\lambda^2+\nothing{\xi}_\text{ow}\lambda+\nothing{\omega}_\text{ow}^2)\frac{1}{\lambda^2}\partial_{\nothing{x}} \nothing{p}|_{\nothing{x}=0}=\nothing{p}(0)\nothing{\rho}
\end{equation}
gives us the following,
\begin{equation}
c_1=c_2\frac{ K_0\left(\frac{e^{2\nothing{k}x_0}}{\nothing{k}\nothing{\sigma}}\sqrt{\lambda^2Z(0,\lambda)}\right) \left(\lambda ^2+\lambda  \nothing{\xi}_{\text{ow}}+\nothing{\omega}_{\text{ow}}^2\right)+ \sqrt{\lambda^2Z(0,\lambda)}\nothing{\rho}  K_1\left(\frac{e^{2\nothing{k}x_0}}{\nothing{k}\nothing{\sigma}}\sqrt{\lambda^2Z(0,\lambda)}\right)}{ I_0\left(\frac{e^{2\nothing{k}x_0}}{\nothing{k}\nothing{\sigma}}\sqrt{\lambda^2Z(0,\lambda)}\right) \left(\lambda ^2+\lambda  \nothing{\xi}_{\text{ow}}+\nothing{\omega}_{\text{ow}}^2\right)-\sqrt{\lambda^2Z(0,\lambda)} \nothing{\rho}  I_1\left(\frac{e^{2\nothing{k}x_0}}{\nothing{k}\nothing{\sigma}}\sqrt{\lambda^2Z(0,\lambda)}\right)}
\end{equation}
Then plugging this definition into the right hand side boundary condition 
\begin{equation}
\label{aRHS}
    \nothing{p}(1)=0
\end{equation}
tells us, 
\small
\begin{equation}
\begin{aligned}    \frac{ K_0\left(\frac{e^{2\nothing{k}x_0}}{\nothing{k}\nothing{\sigma}}\sqrt{\lambda^2Z(0,\lambda)}\right) \left(\lambda ^2+\lambda  \nothing{\xi}_{\text{ow}}+\nothing{\omega}_{\text{ow}}^2\right)+ \sqrt{\lambda^2Z(0,\lambda)}\nothing{\rho}  K_1\left(\frac{e^{2\nothing{k}x_0}}{\nothing{k}\nothing{\sigma}}\sqrt{\lambda^2Z(0,\lambda)}\right)}{ I_0\left(\frac{e^{2\nothing{k}x_0}}{\nothing{k}\nothing{\sigma}}\sqrt{\lambda^2Z(0,\lambda)}\right) \left(\lambda ^2+\lambda  \nothing{\xi}_{\text{ow}}+\nothing{\omega}_{\text{ow}}^2\right)-\sqrt{\lambda^2Z(0,\lambda)} \nothing{\rho}  I_1\left(\frac{e^{2\nothing{k}x_0}}{\nothing{k}\nothing{\sigma}}\sqrt{\lambda^2Z(0,\lambda)}\right)}I_1( \frac{e^{2\nothing{k}x_0}}{\nothing{k}\nothing{\sigma}}\sqrt{\lambda^2Z(1,\lambda)})\\+K_1( \frac{e^{2\nothing{k}x_0}}{\nothing{k}\nothing{\sigma}}\sqrt{\lambda^2Z(1,\lambda)})=0
\end{aligned}
\label{f(lam)=0}
\end{equation}
While we can not solve this directly what we do is use a numerical root solver. In order to find all extended modes as in Fig.~\ref{fig:Extended} we use a line of starting guesses at $-\xi/4+i\omega$ where we change $\omega$ from 0 to $1.2$ times the resonant frequency at the end of the BM. Having a real part of $-\xi/4$ means the starting guesses are halfway between the red line and y-axis.  It is interesting to note however that this procedure on top finding all the extended modes can also find what appear to be solutions left of the red line in Fig. \ref{fig:discont} ($Re[\lambda]<-\frac{\xi}{2}$), at frequencies that would be resonant on the BM.  However when we look at the corresponding eigenfunction we can see that these are not viable solutions of the wave equation as they are discontinuous. The origin of this discontinuity is that $Z(x,\lambda)$ crosses a branch cut in the complex plane at the point at which it is discontinuous. This will occur for all eigenvalues with $Re[\lambda]<-\xi/2$, corresponding to modes lying to the left of the localized-mode band. 
\newpage
\section{Solving for Localized Modes}
In the main text we have shown that extended modes are defined by $Z(x,\lambda)=0$.
 This means for a mode resonant at $x_r$ the eigenvalues are,
 \begin{equation}
     \lambda=-\frac{\nothing{\xi}}{2}\pm i\sqrt{e^{-2\nothing{k}x_0}(1-2\nothing{k}(x_r-x_0))-\frac{\nothing{\xi}^2}{4}}
 \end{equation}
 As in the rest of this work we focus on the positive frequencies. This is the eigenvalue used in the piecewise solution,
 \begin{equation} 
\label{aPW}
\frac{\nothing{p}(\nothing{x},\lambda)}{\sqrt{\lambda^2Z(\nothing{x},\lambda)}}=
 \begin{cases} 
c_{1l}I_1(\frac{e^{2\nothing{k}x_0}}{\nothing{k}\nothing{\sigma}}\sqrt{\lambda^2 Z(\nothing{x},\lambda)})+c_{2l}K_1(\frac{e^{2\nothing{k}x_0}}{\nothing{k}\nothing{\sigma}}\sqrt{\lambda^2 Z(\nothing{x},\lambda)})  &  \nothing{x}< x_r  \\
c_{1r}I_1( \frac{e^{2\nothing{k}x_0}}{\nothing{k}\nothing{\sigma}}\sqrt{\lambda^2 Z(\nothing{x},\lambda)})+c_{2r}K_1( \frac{e^{2\nothing{k}x_0}}{\nothing{k}\nothing{\sigma}}\sqrt{\lambda^2 Z(\nothing{x},\lambda)}) &  \nothing{x}> x_r
\end{cases}
\end{equation}
We begin with the first condition we imposed which is that $\nothing{p}(\nothing{x},\lambda)$ is continuous for $\nothing{x}\in[0,1]$ a fact that is only relevant at $\nothing{x}=x_r$. This means as $\epsilon\rightarrow0^+$,
\begin{equation}
    p(x_r+\epsilon)=p(x_r-\epsilon)
\end{equation}
We will use the expansion of the Bessel functions about 0 to solve this,  
\begin{equation}
    I_1(x)=\frac{x}{2}+O(x^3)~~~~~~~~~K_1(x)=\frac{1}{x}+O(x\ln(x))
\end{equation}
Plugging this in we obtain to leading order,
\begin{equation}
\begin{aligned}
c_{1l}\frac{e^{2\nothing{k}x_0}}{2\nothing{k}\nothing{\sigma}}\sqrt{\lambda^2} e^{-2\nothing{k}x_0}2\nothing{k}\epsilon+c_{2l}(\frac{\nothing{k}\nothing{\sigma}}{e^{2\nothing{k}x_0}}\frac{1}{\sqrt{\lambda^2}})=c_{1r}\frac{e^{2\nothing{k}x_0}}{2\nothing{k}\nothing{\sigma}}\sqrt{\lambda^2} e^{-2\nothing{k}x_0}2\nothing{k}\epsilon+c_{2r}(\frac{\nothing{k}\nothing{\sigma}}{e^{2\nothing{k}x_0}}\frac{1}{\sqrt{\lambda^2}})
\end{aligned}
\end{equation}
Then taking $\epsilon\rightarrow0$ reveals that $c_{2l}=c_{2r}=c_2$
\vspace{0.5cm}

The next condition we impose is the wave equation must hold everywhere, specifically when we integrate
across the resonance,
\begin{equation}
\label{lin_wave}
    \int_{x_r-\epsilon}^{x_r+\epsilon} d\nothing{x}\lambda^2\frac{\nothing{p}(\nothing{x},\lambda)}{\nothing{\sigma} Z(\nothing{x},\lambda)}=    \int_{x_r-\epsilon}^{x_r+\epsilon} d\nothing{x}\partial^2_{\nothing{x}}\nothing{p}(\nothing{x},\lambda)
\end{equation}  
In order to solve this we note that \begin{equation}
    \partial_{\nothing{x}}\nothing{p}=\frac{\lambda^2e^{2\nothing{k}x_0}}{\nothing{k}\nothing{\sigma}}(c_1I_0(\frac{e^{2\nothing{k}x_0}}{\nothing{k}\nothing{\sigma}}\sqrt{\lambda^2 Z(\nothing{x},\lambda)})-c_2K_0(\frac{e^{2\nothing{k}x_0}}{\nothing{k}\nothing{\sigma}}\sqrt{\lambda^2 Z(\nothing{x},\lambda)}))
\end{equation}
\begin{equation}
    \int dx \lambda^2 \frac{p}{Z}=\frac{\lambda^2e^{2\nothing{k}x_0}}{\nothing{k}\nothing{\sigma}}(c_1(1-I_0(\frac{e^{2\nothing{k}x_0}}{\nothing{k}\nothing{\sigma}}\sqrt{\lambda^2 Z(\nothing{x},\lambda)}))+c_2K_0(\frac{e^{2\nothing{k}x_0}}{\nothing{k}\nothing{\sigma}}\sqrt{\lambda^2 Z(\nothing{x},\lambda)}))
\end{equation}
and use the expansion of the zeroth order Bessel functions this time.
\begin{equation}
I_0(x)=1+O(x^2)~~~~~~~~~K_0(x)=\ln(\frac{1}{x})+O(1)
\end{equation}
Then combining these identities and simplifying we get. 
\begin{equation}
    0=c_{1l}-c_{1r}+2c_2(\ln(\sqrt{2\frac{\nothing{k}\nothing{\sigma}e^{-2\nothing{k}x_0}}{-\lambda^2\epsilon}})-\ln(\sqrt{2\frac{\nothing{k}\nothing{\sigma}e^{-2\nothing{k}x_0}}{\lambda^2\epsilon}}))
\end{equation}
In order to evaluate this in the limit that  $\epsilon\rightarrow0$ we use the definition of a complex logarithm $\ln(re^{i\theta})=\ln{r}+i\theta$ restricted to $-\pi<\theta\leq\pi$. Which leads us to, 
\begin{equation}
    c_{1r}-c_{1l}=i\pi c_2
\end{equation} 
These give us the two matching conditions. We then use the boundary condition $\nothing{p}(1)=0$ to get a final constraint. The remaining multiplicative constant is fixed by matching amplitude and phase to the numerical eigenfunction in Fig. \ref{fig:Extended} B. 
\end{document}